\documentclass[twocolumn]{svjour3}          

\usepackage[english]{babel}
\usepackage[T1]{fontenc}
\usepackage[colorlinks=true,linkcolor=blue,citecolor=blue,urlcolor=blue]{hyperref}
\usepackage{bm,bbm,amssymb,amsmath}
\usepackage{graphicx}
\usepackage{multirow,dcolumn,booktabs,enumerate}
\usepackage{isotope}
\usepackage{physics}
\usepackage{slashed}
\usepackage{nicefrac}
\usepackage{xcolor}
\usepackage[normalem]{ulem}

\smartqed  
\begin{document}

\title{Chiral Effective Field Theory's Impact on Advancing Quantum Monte Carlo Methods}

\author{Ingo Tews \and Diego Lonardoni \and Stefano Gandolfi}

\institute{I. Tews \at
              Los Alamos National Laboratory \\
              \email{\href{mailto:itews@lanl.gov}{itews@lanl.gov}}
           \and
           D. Lonardoni \at
              Michigan State University \& Los Alamos National Laboratory \\
              \email{\href{mailto:lonardoni@nscl.msu.edu}{lonardoni@nscl.msu.edu}}
           \and
           S. Gandolfi \at
              Los Alamos National Laboratory \\
              \email{\href{mailto:stefano@lanl.gov}{stefano@lanl.gov}}
}

\date{Received: date / Accepted: date}

\maketitle

\begin{abstract}
Thirty years ago, Steven Weinberg published his seminal paper on ``Nuclear Forces from chiral Lagrangians"~\cite{Weinberg:1990rz} which has revolutionized the field of theoretical nuclear physics. 
Nowadays, interactions derived from chiral effective field theory are routinely used to describe nuclear systems ranging from atomic nuclei to the dense matter explored in the core of neutron stars with theoretical uncertainty estimates.
In our contribution to the special issue ``Celebrating 30 years of the Steven Weinberg’s paper Nuclear Forces from Chiral Lagrangians", we focus on the impact that chiral effective field theory interactions have played in advancing microscopic studies of atomic nuclei and the nuclear-matter equation of state using quantum Monte Carlo methods.\\

\keywords{chiral effective field theory \and quantum Monte Carlo \and nuclear structure \and nuclear dynamics \and nuclear equation of state}
\end{abstract}

\section{Introduction}
\label{intro}

Thirty years ago, Steven Weinberg published a series of papers, starting with ``Nuclear Forces from chiral Lagrangians''~\cite{Weinberg:1990rz} and followed by Refs.~\cite{Weinberg:1991um,Weinberg:1992yk}.
These papers have revolutionized the field of nuclear theory.
Together with Refs.~\cite{Ordonez:1992xp,Ordonez:1993tn,vanKolck:1994yi,Ordonez:1995rz} and the many works that followed them, they established chiral effective field theory (EFT) as a standard tool to derive nuclear forces.
And while there are many open questions still~\cite{Tews:2020hgp}, nowadays interactions from chiral EFT are commonly used in a wide range of applications in nuclear physics, from describing the structure of atomic nuclei to the dense-matter equation of state (EOS) explored in the core of neutron stars~\cite{Epelbaum:2008ga,Epelbaum:2012vx,Hammer:2012id,Hebeler:2015hla,Lynn:2019rdt,Stroberg:2019mxo}.
Most importantly, chiral EFT allows us to estimate theoretical uncertainties originating from the truncation of the input nuclear Hamiltonian~\cite{Epelbaum:2015epja,Drischler:2020yad}. 
This feature sets chiral EFT apart from other microscopic approaches to nuclear forces, and is particularly important when describing nuclear systems in a regime where no experimental data exists. 
For example, analyzing the wealth of recent astrophysical observations of neutron stars (NS) requires knowledge of dense very neutron-rich matter that cannot be probed on Earth.
Such analyses can only be considered reliable if the theoretical uncertainties in the dense matter description can be specified.

The various contributions to the special issue ``Celebrating 30 years of Steven Weinberg’s papers on Nuclear Forces from Chiral Lagrangians'' commemorate several aspects of Steven Weinberg's legacy in nuclear physics~\cite{Witala:2021ufh,Machleidt:2021ggx,Epelbaum:2021sns,Furnstahl:2021rfk,Phillips:2021yet,vanKolck:2021rqu,Drischler:2021kqh}.
In this contribution, we present our personal view of the impact that Weinberg's seminal papers had on microscopic studies of nuclear systems with quantum Monte Carlo (QMC) methods.

Solving the nuclear many-body problem for atomic nuclei and dense nuclear matter is a complicated problem that requires a good understanding of the interactions among the individual nucleons, as well as accurate many-body methods to solve the Schr\"odinger equation.
QMC methods are among the most precise nuclear many-body methods for this task and have been successfully used to describe a variety of nuclear systems and processes~\cite{Carlson:2015}. 
QMC methods solve the Schr\"odinger equation stochastically and, hence, statistical uncertainties can be easily improved by increasing the sample size, \textit{i.e.}, the number of QMC configurations, typically called ``walkers''.
In addition, recent technical developments allow one to obtain quasi-exact results by performing a so-called transient estimation, where systematic uncertainties due to the particular choice of starting wave function for a given system are eliminated~\cite{Lonardoni:2018prc}.
These properties and developments make QMC methods very powerful exact methods to address the nuclear many-body problem. 
In addition, in contrast to other many-body methods, QMC methods are extremely versatile and enable us to use the same Hamiltonians and computational tools to simultaneously study atomic nuclei, nuclear matter, and weak interactions in both.
Hence, QMC methods allow us to truly connect atomic nuclei with the dense matter probed in NSs, transferring knowledge gained in one subfield directly to the other. 
However, an important source of uncertainty remains: \textit{the input nuclear Hamiltonian}.

While QMC methods have been used to describe nuclear systems with great success since the 1990s, they are traditionally combined with phenomenological Hamiltonians describing two-nucleon (NN) and three-nucleon (3N) interactions {\it ad-hoc}. 
Because QMC methods require local coordinate-space input Hamiltonians, they are often used with two-nucleon interactions from the Argonne family~\cite{Wiringa:1995,Wiringa:2002} supplemented with three-nucleon interactions from the Urbana~\cite{Pudliner:1995} or Illinois~\cite{Pieper:2001} families.
These interactions are very successful when describing nuclear systems~\cite{Carlson:2015} but have several shortcomings: 
most importantly, theoretical uncertainties cannot be specified and it is unclear how nuclear forces, in particular 3N forces, can be systematically improved.

Weinberg's papers on nuclear forces have revolutionized the field.
Most, if not all current calculations of nuclear systems that start from microscopic Hamiltonians, employ chiral EFT interactions.
While chiral EFT interactions conquered most of nuclear physics, it took more than 20 years for these interactions to be used in QMC computational methods. 
The combination of these two important tools has enabled many exciting studies leading to fascinating insights and results and realizing the full potential of QMC methods~\cite{Lynn:2019rdt}.
In this contribution, we discuss the impact of Weinberg's proposal on advancing QMC methods and review some exciting recent results spanning from nuclei to nuclear matter explored in NSs.

\section{QMC and chiral EFT}
\label{QMC_EFT}

\subsection{A brief review of QMC methods}

Quantum Monte Carlo methods refer to a family of stochastic techniques developed to solve the many-body ground state of correlated systems~\cite{Foulkes:2001}. 
For strongly correlated systems, such as nuclear systems, variational and diffusion Monte Carlo methods are widely used~\cite{Carlson:2015}. 
The latter in particular rely on the projection of an initial trial wave function, $\Psi_T$, that has the quantum numbers representing the system, into the ground-state wave function $\Psi_0$ via the imaginary time propagation:
\begin{equation}
    e^{-(H-E_T)\tau}\Psi_T(R,S)\stackrel{\tau\to\infty}{\longrightarrow}\Psi_0(R',S')\,,
\end{equation}
where $R=\{r_1,\ldots,r_A\}$ and $S=\{s_1,\ldots,s_A\}$ represent the coordinates and spin-isospin projections of the $A$ nucleons, respectively, $H$ is the Hamiltonian of the system, $E_T$ is a parameter that controls the normalization, and the long imaginary time propagation is carried out by applying a large number of small time-step iterations. 
This procedure is implemented by considering a collection of coordinate/spin/isospin configurations (called walkers) that are simultaneously evolved in imaginary time. 
Nucleon positions are diffused $(R\to R')$ according the the kinetic energy of the Hamiltonian, while spin/isospin configurations are modified $(S\to S')$ according to the spin/isospin-dependent part of the nuclear potential. 
A weight is associated to each new configuration and it is used as a survival probability, \textit{i.e.}, a branching prescription is applied (branching random walk). 

Expectation values of the quantity of interest are evaluated on the propagated walkers. 
The so-called sign problem that affects most diffusion Monte Carlo techniques, is kept under control by implementing different prescriptions to prevent walkers from crossing the multi-dimensional nodal surface determined by the employed wave function~\cite{Carlson:2015}. 
Depending on the quality of the input wave function, this might lead to biases in the ground-state expectation values. 
Such biases are removed by performing an unconstrained propagation (transient estimation) after the constrained propagation is converged.
This transient estimation is performed for a relatively small imaginary-time window before the sign problem manifests itself. 
For a good input wave function, the system converges to the true ground-state value within this window~\cite{Lonardoni:2018prc}. 
This eliminates the systematic uncertainties due to the particular choice of starting wave function and, hence, produces statistically exact results, making QMC methods extremely powerful.

Green's function Monte Carlo (GFMC)~\cite{Carlson:1987} and auxiliary field diffusion Monte Carlo (AFDMC)~\cite{Schmidt:1999} are two well established diffusion Monte Carlo techniques for nuclear systems. 
The former retains the full spin/isospin structure of the wave function during the imaginary time propagation. 
The latter employs a somewhat simplified wave function, and spin/isospin rotations are instead sampled during the propagation. 
GFMC is best suited for relatively small nuclei $(A\lesssim12)$, due to the exponential scaling of the method with the number of particles. 
The advantage of AFDMC is an overall reduced computational cost that allows one to study larger systems, from nuclei up to $A\approx 40$ nucleons~\cite{Piarulli:2017dwd,Lonardoni:2018prl,Lonardoni:2018prc,Lonardoni:2018rhok,Lynn:2020,Cruz:2021}, to nuclear matter~\cite{Lynn:2015jua,Tews:2018kmu,Piarulli:2019pfq,Lonardoni:2020}.
This makes the AFDMC method an ideal tool to study properties for a range of nuclear systems, using the same theoretical input, and allows us to test nuclear Hamiltonians against laboratory data and astrophysical observations of NSs.

\subsection{Using chiral EFT interactions in QMC methods}

Due to the sign problem, the applicability of both GFMC and AFDMC methods is limited to nearly local Hamiltonians only, \textit{i.e.}, Hamiltonians that only depend on the relative distance $\textbf{r}$ between individual nucleons and can be formulated in coordinate space.
Historically, this drove the development and vast application of phenomenological two-~\cite{Wiringa:1995,Wiringa:2002} and three-body forces~\cite{Pudliner:1995,Pieper:2001}.

All calculations of nuclear systems suffer from two sources of uncertainty: the theoretical uncertainty of the nuclear interaction used to describe a system, and the method uncertainty introduced by assumptions and approximations made to solve the nuclear many-body problem. 
The latter is extremely small for QMC methods~\cite{Carlson:2015}, making them ideal computational tools to study various nuclear systems.
Unfortunately, it is not possible to quantify theoretical uncertainties for commonly used phenomenological interactions because they are not based on a systematic framework. 
The advent of chiral EFT interactions enabled the exciting opportunity to estimate theoretical uncertainties in calculations of nuclear systems.
Initially, these estimates were obtained by varying certain low-energy couplings (LECs) or the cutoff scale of nuclear Hamiltonians~\cite{Hebeler:2009iv,Tews:2012fj,Drischler:2013iza,Simonis:2015vja}, but in the past years improved descriptions to estimate theoretical uncertainties~\cite{Epelbaum:2015epja} were widely used~\cite{Lynn:2015jua,Drischler:2017wtt,Lonardoni:2018prl}.
Recently, new computational tools for error estimation using Gaussian processes became available~\cite{Drischler:2020yad}.
Unfortunately, for a long time, chiral EFT interactions could not be implemented in QMC methods.

As written above, QMC methods require local input Hamiltonians.
Chiral EFT, however, is naturally formulated in momentum space and contains nonlocal terms. 
In particular, chiral EFT NN interactions depend on two momentum scales: the momentum transfer between two nucleons $\textbf{q}$ and the momentum transfer in the exchange channel $\textbf{k}$. 
Upon Fourier transformation to coordinate space, the first one results in local interactions depending on $\textbf{r}$, while the latter leads to nonlocalities in the form of derivatives that act on the wave function.
Unfortunately, non-local interactions are very difficult to be propagated in GFMC and AFDMC, and only very small terms can be estimated using perturbation theory.
That is not the case of most of the non-local chiral EFT potentials.
Hence, state-of-the-art chiral EFT interactions could not be used in QMC methods for a long time.

This changed in 2013, when the first local chiral EFT interactions were implemented in QMC methods~\cite{Gezerlis:2013}. 
Based on the initial work of Ref.~\cite{Freunek:61123}, the idea is to use the Fierz-rearrangement freedom of the short-range part of chiral EFT interactions to replace all nonlocal terms in the nuclear NN interactions with local ones~\cite{Gezerlis:2014,Huth:2017wzw,Piarulli:2019cqu}.
This can be achieved because among all possible interaction terms in the chiral potential, due to the antisymmetry of nuclear wave functions, only half of them are linearly independent. 
For example, at leading order (LO) there are four general momentum-independent spin-isopin interactions:
\begin{equation}
    V_{\rm LO}=\alpha_1\,\mathbbm{1}+\alpha_{\sigma\,}\sigma_{ij}+\alpha_{\tau}\,\tau_{ij}+\alpha_{\sigma\tau}\,\sigma_{ij}\,\tau_{ij}\,,
\end{equation}
where $\sigma_{ij}={\bm\sigma}_i\cdot{\bm\sigma}_j$ and similarly for $\tau_{ij}$, and the $\alpha_i$ are LECs.
Only two out of these four interactions are linearly independent, describing the two $S$-wave interaction channels.
Similarly at next-to-leading order (NLO), 7 out of 14 interaction terms have to be chosen, and it is possible to select a set of 6 local terms and the nonlocal spin-orbit term, which can however be treated in QMC methods.
In addition, the leading 3N interactions at next-to-next-to-leading order (N$^2$LO) are local and can be implemented in QMC methods~\cite{Tews:2015ufa,Lynn:2015jua}.
Hence, it is possible to construct local chiral coordinate-space interactions up to N$^2$LO ~\cite{Gezerlis:2013,Gezerlis:2014}.

Since this initial work, additional local chiral EFT interactions have been constructed, including selected local NN interaction terms at next-to-next-to-next-to-leading order (N$^3$LO) and also including Delta-isobars~\cite{Piarulli:2015,Piarulli:2016,Piarulli:2018prl,Piarulli:2018}.
Both families of local chiral EFT interactions have opened the exciting opportunity to combine chiral EFT potentials and quasi-exact quantum Monte Carlo calculations and lead to many exciting results~\cite{Lynn:2015jua,Piarulli:2016,Lonardoni:2018prl,Lonardoni:2020,King:2020,Cruz:2021,Schiavilla:2021}.
An important benefit of QMC is that it can treat bare chiral EFT interactions also at large cutoff values, that are usually too ``stiff'' for other many-body methods. 
Hence, this combination allows us to address high-cutoff interactions and enables us to study fundamental problems of EFT interactions, \textit{e.g.}, the appearance of spurious bound states~\cite{Tews:2018sbi}.
In the following, we will present some recent highlights of $\Delta$-less chiral EFT potentials up to N$^2$LO that have been used in AFDMC studies of atomic nuclei and dense nuclear matter. 

\section{Recent highlights}
\label{Highlights}

The advent of local chiral EFT potentials formulated in coordinate space allowed for the first diffusion Monte Carlo calculations of nuclei beyond $A=12$~\cite{Lonardoni:2018prl}, of the equation of state of nuclear matter and the symmetry energy~\cite{Lonardoni:2020}, and for novel QMC calculations of the neutron-matter EOS~\cite{Lynn:2015jua,Tews:2018kmu} with uncertainty estimates. 
Below, we present recent highlights from these and follow-up studies.

\subsection{Nuclei and short-range correlations}

The AFDMC method in combination with chiral EFT potentials has been used to calculate ground-state properties of nuclei up to \isotope[16]{O}. 
It has been shown that $\Delta$-less local chiral EFT interactions at N$^2$LO, that are only fit to NN phase shifts and properties of light systems, are capable of simultaneously reproducing both binding energies and charge radii of nuclei (at least) up to $A=16$~\cite{Lonardoni:2018prl,Lonardoni:2018prc}, as well as neutron-\isotope[4]{He} scattering phase shifts~\cite{Lynn:2015jua,Lonardoni:2018prc}.
Many other properties have been also derived, from densities~\cite{Gandolfi:2020} to charge form factors~\cite{Lonardoni:2018prc} to momentum distributions~\cite{Lonardoni:2018rhok}, showing very good agreement with experimental data where available.
For most of these quantities, theoretical uncertainties due to the truncation of the chiral expansion are specified.

Of particular interest is the application of QMC methods with chiral interactions to the study of nuclear short-range correlations (SRCs). 
In strongly interacting Fermion systems, such as nuclear systems, the short-range component of the interaction is responsible for the creation of short-range-correlated pairs, pairs of Fermions with a large relative momentum $(q > k_F)$ and a small center-of-mass momentum $(K < k_F)$, where $k_F$ is the Fermi momentum of the system.
In atomic nuclei, short-range-correlated pairs have been studied using many different reactions, including pick-up, stripping, and electron and proton scattering. 
The results of these and many other studies highlighted the importance of correlations in nuclei. 
SRCs have substantial implications for nuclear structure properties, for the internal structure of nucleons bound in nuclei, for neutrinoless double beta decay matrix elements, for the nuclear equation of state, and for properties of neutron stars~\cite{Hen:2017}. 
Therefore, understanding their formation mechanisms and specific characteristics is required for obtaining a complete description of nuclear systems. 

\begin{figure}[t]
    \includegraphics[width=0.98\linewidth]{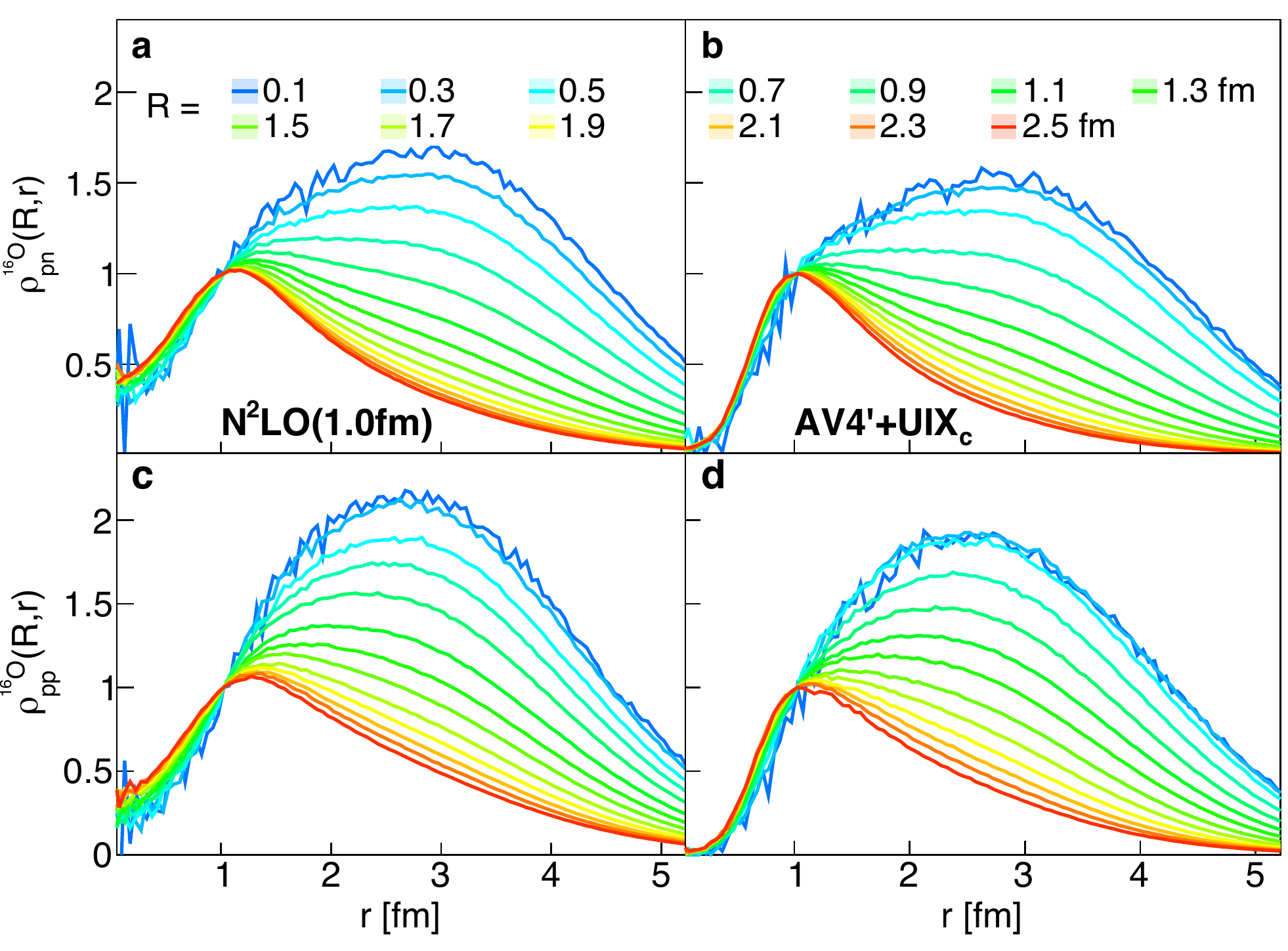}
    \caption[]{Short-distance universality of two-nucleon density in \isotope[16]{O}. The densities $\rho(R,r)$ are shown as a function of the relative-distance $r$ for different pair center-of-mass position $R$ in the nucleus. (a,b) for $pn$ pairs, (c,d) for $pp$ pairs. (a,c) for a local chiral interaction, (b,d) for a phenomenological potential. See Ref.~\cite{Cruz:2021} for details.}
    \label{fig:rho2_o16}
\end{figure}

In a recent work published in {\it Nature Physics}~\cite{Cruz:2021}, two-body distributions in both coordinate and momentum space derived from QMC calculations using different nuclear potentials, including those derived from chiral EFT, have been used to quantitatively study SRC effects in nuclei from deuterium to \isotope[40]{Ca}. 
The study reports a universal factorization of the many-body nuclear wave function at short distance into a strongly interacting pair and a weakly interacting residual system (see Fig.~\ref{fig:rho2_o16}). 
The residual system distribution is consistent with that of an uncorrelated system, showing that short-distance correlation effects are predominantly embedded in two-body correlations.
Moreover, from the extraction of spin- and isospin-dependent ``nuclear contact terms'', the position-momentum equivalence of nuclear SRCs has been proven. 
Finally, the contact coefficient ratio between two different nuclei shows very little dependence on the nuclear interaction model. 
These novel findings allow one to extend the application of mean-field approximations to short-range correlated pair formation by showing that the relative abundance of short-range pairs in the nucleus is a long-range (that is, mean field) quantity that is insensitive to the short-distance nature of the nuclear force. 

As highlighted by this recent work, the combination of QMC methods and chiral EFT potentials provides a new and versatile tool to study the somewhat elusive nuclear SRCs.

\subsection{The equation of state of neutron stars and multimessenger constraints}

\begin{figure}[t]
    \includegraphics[width=0.98\linewidth]{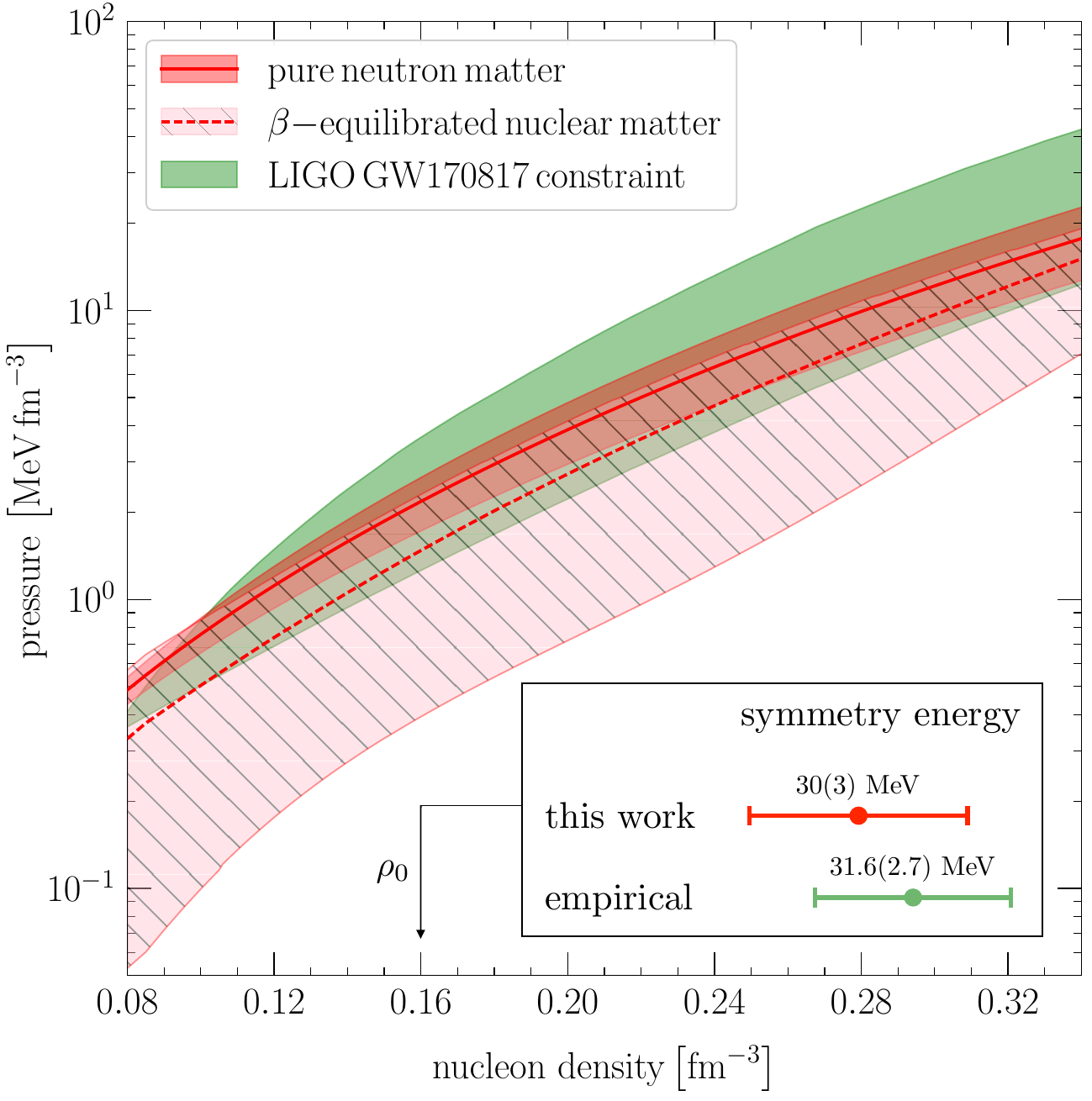}
    \caption[]{Equation of state of pure neutron matter (red band with solid line) and $\beta$-equilibrated nuclear matter (red hatched band with dashed line) for local chiral interactions. The green area is the pressure extracted from the gravitational wave signal GW170817 for EOSs reproducing 2 solar-mass neutron stars~\cite{Abbott:2018}. The inset reports the extracted and empirical values of the symmetry energy at saturation density $\rho_0$. Adapted from Ref.~\cite{Lonardoni:2020}.}
    \label{fig:eos}
\end{figure}

The same AFDMC method and the same interactions have also been used to study nucleonic matter, ranging from pure neutron matter relevant to describe neutron stars~\cite{Lynn:2015jua,Tews:2018kmu} to symmetric nuclear matter and the symmetry energy~\cite{Lonardoni:2020} that can be probed in heavy-ion collision experiments.
It was shown in these studies that the same approach that was successful for atomic nuclei also provides reliable description of dense matter. 
We show this in Fig.~\ref{fig:eos}, where the calculated pressure is shown as a function of nucleon number density for pure neutron matter and matter in $\beta$-equilibrium.
For the latter, the calculated symmetry energy (and no empirical information) was used to extract the proton fraction in beta equilibrium.
We compare the QMC results with an extraction of the pressure of $\beta$-equilibrated matter by the LIGO collaboration, using gravitational-wave data from the signal GW170817~\cite{Abbott:2018}.
In addition, we compare the calculated symmetry energy with the empirical value for the symmetry energy from Ref.~\cite{Li:2019}.

\begin{figure*}[t]
    \centering
    \includegraphics[width=0.98\textwidth]{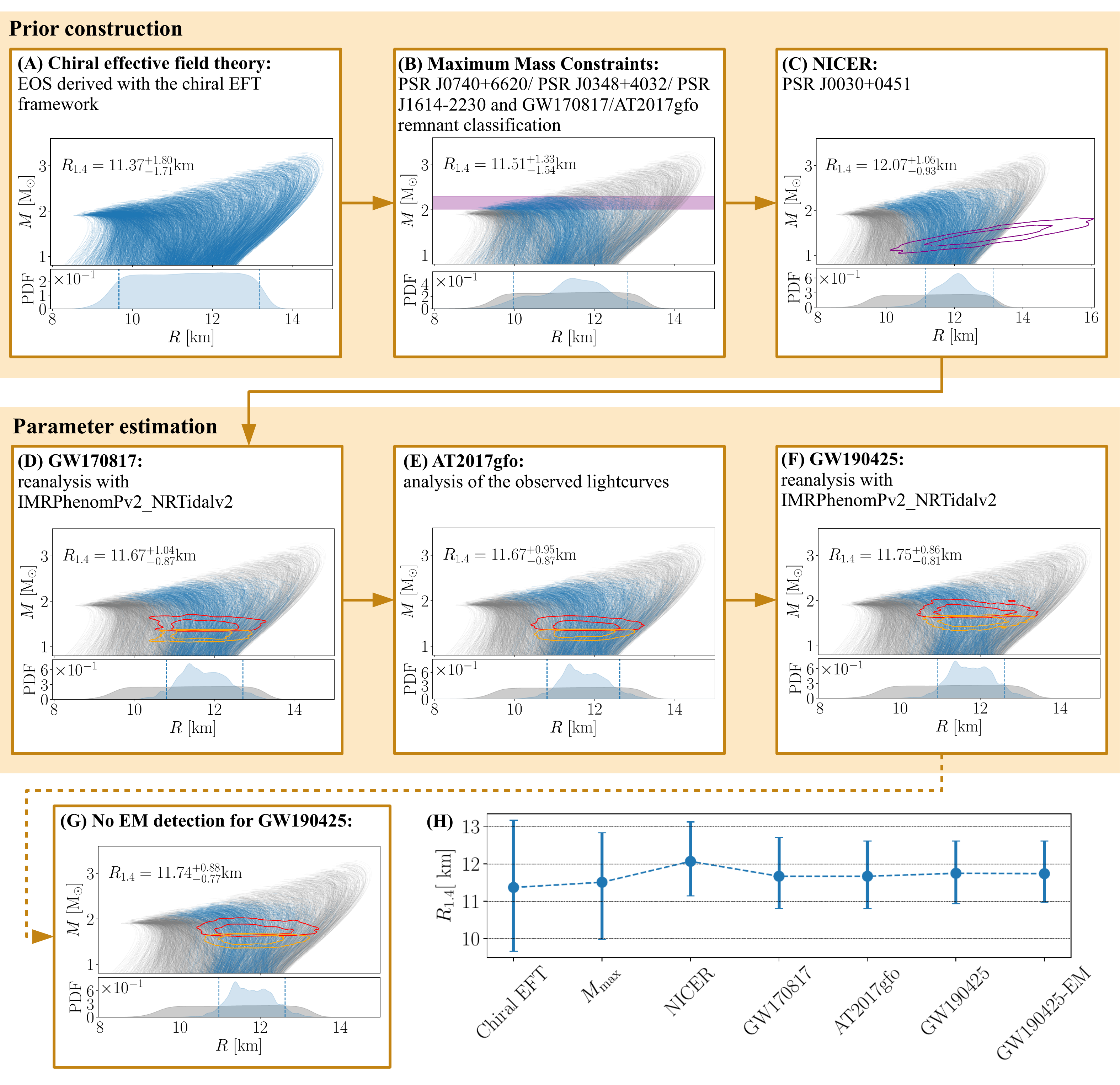}
    \caption{Constraints on the NS EOS when analyzing different astrophysical observations in a multi-step procedure~\cite{Dietrich:2020efo}.
    Allowed (disallowed) EOS are shown as blue (gray) lines in each step and the lower panels show the probability distribution function (PDF) for the radius of a 1.4 solar mass NS with the 90\% confidence range indicated by dashed lines.
    The analysis begins with an EOS set constrained by QMC calculations using chiral EFT interactions (A).
    These EOS are then constrained by information on the maximum NS mass from pulsar mass measurements and the maximum-mass constraints obtained from GW170817 and its EM counterpart AT2017gfo (B), NICER mass-radius measurements of PSR~J0030+0451 (C), GW170817 (D), AT2017gfo (E), GW190425 (F) and the assumption that GW190425 did not produce a detectable EM signal (G).
    The radius constraint at 90\% confidence is summarized for each analysis step in panel (H). Figure taken from Ref.~\cite{Dietrich:2020efo}.}
    \label{fig:scheme}
\end{figure*}

The excellent agreement of the QMC calculations and the experimental/observational extractions of dense-matter properties shows the reliability of QMC calculations with local chiral EFT interactions.
This is extremely important when studying recent observational constraints on NSs, where these calculations have to be extrapolated to very neutron-rich conditions and high densities.
Only for reliable nuclear-physics input can these observations be analyzed with confidence. 
For this, the possibility to specify theoretical uncertainties is crucial.
In Fig.~\ref{fig:scheme}, we show the results of a recent study presented in {\it Science}~\cite{Dietrich:2020efo}.
In this study, a wealth of recent NS multimessenger observations has been analyzed using Bayesian inference for the NS EOS. 
This analysis starts from a set of NS EOS that was constrained below 1.5 times the nuclear saturation density $\rho_0$ by the QMC calculations discussed above and extended to higher densities using a general speed-of-sound extension scheme~\cite{Tews:2018iwm}.
Such general schemes are ideal to extend low-density uncertainty estimates to higher densities without making any assumption about the high-density EOS, and thereby, not inducing strong systematic uncertainties.
The only constraints that are typically implemented are causality (speed of sound is smaller than the speed of light) and stability (speed of sound is not negative).
By analyzing information on the maximum mass of neutron stars from pulsar timing observations~\cite{Demorest:2010bx,Antoniadis:2013pzd,Arzoumanian:2017puf,Cromartie:2019kug}, 
$X$-ray pulse-profile modeling of the pulsar J0030+0451 by the Neutron-Star Interior Composition Explorer (NICER)~\cite{Miller:2019cac,Riley:2019yda},
gravitational-wave observations of binary NS mergers, GW170817~\cite{Abbott:2018} and GW190425~\cite{LIGOScientific:2020aai}, the electromagnetic (EM) counterparts associated with GW170817, AT2017gfo and GRB170817A~\cite{GBM:2017lvd,Arcavi:2017xiz,Coulter:2017wya}, and maximum mass inferences from these counterparts~\cite{Rezzolla:2017aly}, Ref.~\cite{Dietrich:2020efo} provided most stringent constraints on the radii of NSs and the Hubble constant.

These Bayesian multimessenger analyses can easily include additional information from NS observations~\cite{Tews:2020ylw,Al-Mamun:2020vzu,Pang:2021jta}, and nuclear experiments, \textit{e.g.}, heavy-ion collision experiments~\cite{Huth:2021bsp} or the recent PREX experiment~\cite{Essick:2021kjb,Essick:2021ezp} from which the pressure among neutrons can be extracted~\cite{PREXII,Reed:2021nqk}. 
With more and more NS observations that are expected in the coming years, constraints on the EOS will become tighter.
This will allow us to probe the agreement of chiral EFT prediction with observational data to estimate the breakdown density of chiral EFT in dense neutron-star matter~\cite{Essick:2020flb} and to constrain nuclear Hamiltonians directly with astrophysical data.
For this, calculations with estimates of the theoretical uncertainty due to nuclear Hamiltonians are key.
This would not be possible without Steven Weinberg's seminal work.

\section{Summary}
\label{summary}

One of the keys for the reliability of studies of atomic nuclei and dense matter is the possibility to employ the same Hamiltonians and many-body methods over a wide range of physical systems. 
Here, we have discussed the combination of QMC methods with local interactions from chiral EFT as a versatile tool to systematically study all of these systems.
The calculations of atomic nuclei and their various properties have shown excellent agreement with experimental data, while studies of dense matter agree very well with empirical knowledge and astrophysical observations. 

Robust analyses of these systems require a way of estimating the uncertainties of the calculations.
The small statistical uncertainties of QMC methods make them ideal many-body methods for this purpose, but they could only reach their full potential due to the combination with interactions from chiral EFT. 
Analyses of NS observations in particular rely sensitively on nuclear-theory input with uncertainty estimates that make calculations reliable and falsifiable.
Without such estimates, which were only made possible through Weinberg's seminal papers, the analysis of exciting present and future observations of NSs by Ligo or NICER would not be robust.

The ground-breaking work of Steven Weinberg from 30 years ago revolutionized the field of nuclear physics. 
It lead to exciting results for atomic nuclei, from nuclear structure to nuclear dynamics.
As it turns out, it also revolutionized the field of NS astrophysics.
Chiral EFT calculations have become one of the important inputs in current analyses of exciting new astrophyscial observations~\cite{Most:2018hfd,Raaijmakers:2019qny,Capano:2019eae,Dietrich:2020efo}.

A lot of work still has to be done~\cite{Tews:2020hgp}. 
For example, while nonlocal nuclear interactions in the NN sector have been constructed up to N$^4$LO~\cite{Entem:2017gor,Reinert:2017usi}, the consistent derivation of three- and many-body forces is still underway, and it will pose a great challenge to all the many-body methods.
Also, differences in regulator choices, in particular local vs non-local regulators, have to be better understood.
Nevertheless, the progresses in deriving nuclear interactions that have been made in recent years by the community opened the way to revolutionize studies of nuclear systems.
With this, we are entering the time of precision nuclear physics.

\begin{acknowledgements}
The work of I.T. was supported by the U.S. Department of Energy, Office of Science, Office of Nuclear Physics, under contract No.~DE-AC52-06NA25396, by the Laboratory Directed Research and Development program of Los Alamos National Laboratory under project numbers 20190617PRD1 and 20190021DR, and by the U.S. Department of Energy, Office of Science, Office of Advanced Scientific Computing Research, Scientific Discovery through Advanced Computing (SciDAC) NUCLEI program.
The work of D.L. was supported by the U.S. Department of Energy, Office of Science, Office of Nuclear Physics, under Contract No.~DE-SC0013617, and by the NUCLEI SciDAC program. 
The work of S.G. was supported by U.S.~Department of Energy, Office of Science,  Office of Nuclear Physics, under Contract No.~DE-AC52-06NA25396, by the DOE NUCLEI SciDAC Program, by the LANL LDRD Program, and by the DOE Early Career Research Program.
Computational resources have been provided  by the Los Alamos National Laboratory Institutional Computing Program, which is supported by the U.S. Department of Energy National Nuclear Security Administration under Contract No.~89233218CNA000001, and by the National Energy Research Scientific Computing Center (NERSC), which is supported by the U.S. Department of Energy, Office of Science, under contract No.~DE-AC02-05CH11231.
\end{acknowledgements}

\bibliographystyle{spphys}       
\bibliography{chiral}

\begin{thebibliography}{10}
\providecommand{\url}[1]{{#1}}
\providecommand{\urlprefix}{URL }
\expandafter\ifx\csname urlstyle\endcsname\relax
  \providecommand{\doi}[1]{DOI \discretionary{}{}{}#1}\else
  \providecommand{\doi}{DOI \discretionary{}{}{}\begingroup
  \urlstyle{rm}\Url}\fi

\bibitem{Weinberg:1990rz}
S.~Weinberg, Phys. Lett. B \textbf{251}, 288 (1990).
\newblock \doi{10.1016/0370-2693(90)90938-3}

\bibitem{Weinberg:1991um}
S.~Weinberg, Nucl. Phys. B \textbf{363}, 3 (1991).
\newblock \doi{10.1016/0550-3213(91)90231-L}

\bibitem{Weinberg:1992yk}
S.~Weinberg, Phys. Lett. B \textbf{295}, 114 (1992).
\newblock \doi{10.1016/0370-2693(92)90099-P}

\bibitem{Ordonez:1992xp}
C.~Ordonez, U.~van Kolck, Phys. Lett. B \textbf{291}, 459 (1992).
\newblock \doi{10.1016/0370-2693(92)91404-W}

\bibitem{Ordonez:1993tn}
C.~Ordonez, L.~Ray, U.~van Kolck, Phys. Rev. Lett. \textbf{72}, 1982 (1994).
\newblock \doi{10.1103/PhysRevLett.72.1982}

\bibitem{vanKolck:1994yi}
U.~van Kolck, Phys. Rev. C \textbf{49}, 2932 (1994).
\newblock \doi{10.1103/PhysRevC.49.2932}

\bibitem{Ordonez:1995rz}
C.~Ordonez, L.~Ray, U.~van Kolck, Phys. Rev. C \textbf{53}, 2086 (1996).
\newblock \doi{10.1103/PhysRevC.53.2086}

\bibitem{Tews:2020hgp}
I.~Tews, Z.~Davoudi, A.~Ekstr\"om, J.D. Holt, J.E. Lynn, J. Phys. G
  \textbf{47}(10), 103001 (2020).
\newblock \doi{10.1088/1361-6471/ab9079}

\bibitem{Epelbaum:2008ga}
E.~Epelbaum, H.W. Hammer, U.G. Meissner, Rev. Mod. Phys. \textbf{81}, 1773
  (2009).
\newblock \doi{10.1103/RevModPhys.81.1773}

\bibitem{Epelbaum:2012vx}
E.~Epelbaum, U.G. Meissner, Ann. Rev. Nucl. Part. Sci. \textbf{62}, 159 (2012).
\newblock \doi{10.1146/annurev-nucl-102010-130056}

\bibitem{Hammer:2012id}
H.W. Hammer, A.~Nogga, A.~Schwenk, Rev. Mod. Phys. \textbf{85}, 197 (2013).
\newblock \doi{10.1103/RevModPhys.85.197}

\bibitem{Hebeler:2015hla}
K.~Hebeler, J.D. Holt, J.~Menendez, A.~Schwenk, Ann. Rev. Nucl. Part. Sci.
  \textbf{65}, 457 (2015).
\newblock \doi{10.1146/annurev-nucl-102313-025446}

\bibitem{Lynn:2019rdt}
J.E. Lynn, I.~Tews, S.~Gandolfi, A.~Lovato, Ann. Rev. Nucl. Part. Sci.
  \textbf{69}, 279 (2019).
\newblock \doi{10.1146/annurev-nucl-101918-023600}

\bibitem{Stroberg:2019mxo}
S.R. Stroberg, S.K. Bogner, H.~Hergert, J.D. Holt, Ann. Rev. Nucl. Part. Sci.
  \textbf{69}, 307 (2019).
\newblock \doi{10.1146/annurev-nucl-101917-021120}

\bibitem{Epelbaum:2015epja}
E.~Epelbaum, H.~Krebs, U.G. Mei\ss{}ner, Eur. Phys. J. A \textbf{51}(5), 53
  (2015).
\newblock \doi{10.1140/epja/i2015-15053-8}

\bibitem{Drischler:2020yad}
C.~Drischler, J.A. Melendez, R.J. Furnstahl, D.R. Phillips, Phys. Rev. C
  \textbf{102}(5), 054315 (2020).
\newblock \doi{10.1103/PhysRevC.102.054315}

\bibitem{Witala:2021ufh}
H.~Wita\l{}a, J.~Golak, R.~Skibi\'nski, K.~Topolnicki, Few Body Syst.
  \textbf{62}(2), 23 (2021).
\newblock \doi{10.1007/s00601-021-01607-2}

\bibitem{Machleidt:2021ggx}
R.~Machleidt, Few Body Syst. \textbf{62}(2), 21 (2021).
\newblock \doi{10.1007/s00601-021-01606-3}

\bibitem{Epelbaum:2021sns}
E.~Epelbaum, J.~Gegelia, H.P. Huesmann, U.G. Mei\ss{}ner, X.L. Ren, Few Body
  Syst. \textbf{62}(3), 51 (2021).
\newblock \doi{10.1007/s00601-021-01628-x}

\bibitem{Furnstahl:2021rfk}
R.J. Furnstahl, H.W. Hammer, A.~Schwenk, Few Body Syst. \textbf{62}(3), 72
  (2021).
\newblock \doi{10.1007/s00601-021-01658-5}

\bibitem{Phillips:2021yet}
D.R. Phillips, arXiv:2107.03558  (2021)

\bibitem{vanKolck:2021rqu}
U.~van Kolck, arXiv:2107.11675  (2021)

\bibitem{Drischler:2021kqh}
C.~Drischler, S.K. Bogner, arXiv:2108.03771  (2021)

\bibitem{Carlson:2015}
J.~Carlson, S.~Gandolfi, F.~Pederiva, S.C. Pieper, R.~Schiavilla, K.E. Schmidt,
  R.B. Wiringa, Rev. Mod. Phys. \textbf{87}, 1067 (2015).
\newblock \doi{10.1103/RevModPhys.87.1067}

\bibitem{Lonardoni:2018prc}
D.~Lonardoni, S.~Gandolfi, J.E. Lynn, C.~Petrie, J.~Carlson, K.E. Schmidt,
  A.~Schwenk, Phys. Rev. C \textbf{97}, 044318 (2018).
\newblock \doi{10.1103/PhysRevC.97.044318}

\bibitem{Wiringa:1995}
R.B. Wiringa, V.G.J. Stoks, R.~Schiavilla, Phys. Rev. C \textbf{51}, 38 (1995).
\newblock \doi{10.1103/PhysRevC.51.38}

\bibitem{Wiringa:2002}
R.B. Wiringa, S.C. Pieper, Phys. Rev. Lett. \textbf{89}, 182501 (2002).
\newblock \doi{10.1103/PhysRevLett.89.182501}

\bibitem{Pudliner:1995}
B.S. Pudliner, V.R. Pandharipande, J.~Carlson, R.B. Wiringa, Phys. Rev. Lett.
  \textbf{74}, 4396 (1995).
\newblock \doi{10.1103/PhysRevLett.74.4396}

\bibitem{Pieper:2001}
S.C. Pieper, V.R. Pandharipande, R.B. Wiringa, J.~Carlson, Phys. Rev. C
  \textbf{64}, 014001 (2001).
\newblock \doi{10.1103/PhysRevC.64.014001}

\bibitem{Foulkes:2001}
W.M.C. Foulkes, L.~Mitas, R.J. Needs, G.~Rajagopal, Rev. Mod. Phys.
  \textbf{73}, 33 (2001).
\newblock \doi{10.1103/RevModPhys.73.33}

\bibitem{Carlson:1987}
J.~Carlson, Phys. Rev. C \textbf{36}(5), 2026 (1987).
\newblock \doi{10.1103/PhysRevC.36.2026}

\bibitem{Schmidt:1999}
K.E. Schmidt, S.~Fantoni, Phys. Lett. B \textbf{446}(2), 99 (1999).
\newblock \doi{10.1016/S0370-2693(98)01522-6}

\bibitem{Piarulli:2017dwd}
M.~Piarulli, et~al., Phys. Rev. Lett. \textbf{120}(5), 052503 (2018).
\newblock \doi{10.1103/PhysRevLett.120.052503}

\bibitem{Lonardoni:2018prl}
D.~Lonardoni, J.~Carlson, S.~Gandolfi, J.E. Lynn, K.E. Schmidt, A.~Schwenk,
  X.B. Wang, Phys. Rev. Lett. \textbf{120}, 122502 (2018).
\newblock \doi{10.1103/PhysRevLett.120.122502}

\bibitem{Lonardoni:2018rhok}
D.~Lonardoni, S.~Gandolfi, X.B. Wang, J.~Carlson, Phys. Rev. C \textbf{98},
  014322 (2018).
\newblock \doi{10.1103/PhysRevC.98.014322}

\bibitem{Lynn:2020}
J.E. Lynn, D.~Lonardoni, J.~Carlson, J.W. Chen, W.~Detmold, S.~Gandolfi,
  A.~Schwenk, J. Phys. G: Nucl. Part. Phys. \textbf{47}(4), 045109 (2020).
\newblock \doi{10.1088/1361-6471/ab6af7}

\bibitem{Cruz:2021}
R.~Cruz-Torres, D.~Lonardoni, R.~Weiss, M.~Piarulli, N.~Barnea, D.W.
  Higinbotham, E.~Piasetzky, A.~Schmidt, L.B. Weinstein, R.B. Wiringa, O.~Hen,
  Nature Physics \textbf{17}, 306 (2021).
\newblock \doi{10.1038/s41567-020-01053-7}

\bibitem{Lynn:2015jua}
J.E. Lynn, I.~Tews, J.~Carlson, S.~Gandolfi, A.~Gezerlis, K.E. Schmidt,
  A.~Schwenk, Phys. Rev. Lett. \textbf{116}(6), 062501 (2016).
\newblock \doi{10.1103/PhysRevLett.116.062501}

\bibitem{Tews:2018kmu}
I.~Tews, J.~Carlson, S.~Gandolfi, S.~Reddy, Astrophys. J. \textbf{860}(2), 149
  (2018).
\newblock \doi{10.3847/1538-4357/aac267}

\bibitem{Piarulli:2019pfq}
M.~Piarulli, I.~Bombaci, D.~Logoteta, A.~Lovato, R.B. Wiringa, Phys. Rev. C
  \textbf{101}(4), 045801 (2020).
\newblock \doi{10.1103/PhysRevC.101.045801}

\bibitem{Lonardoni:2020}
D.~Lonardoni, I.~Tews, S.~Gandolfi, J.~Carlson, Phys. Rev. Research \textbf{2},
  022033 (2020).
\newblock \doi{10.1103/PhysRevResearch.2.022033}

\bibitem{Hebeler:2009iv}
K.~Hebeler, A.~Schwenk, Phys. Rev. C \textbf{82}, 014314 (2010).
\newblock \doi{10.1103/PhysRevC.82.014314}

\bibitem{Tews:2012fj}
I.~Tews, T.~Kr\"uger, K.~Hebeler, A.~Schwenk, Phys. Rev. Lett. \textbf{110}(3),
  032504 (2013).
\newblock \doi{10.1103/PhysRevLett.110.032504}

\bibitem{Drischler:2013iza}
C.~Drischler, V.~Soma, A.~Schwenk, Phys. Rev. C \textbf{89}(2), 025806 (2014).
\newblock \doi{10.1103/PhysRevC.89.025806}

\bibitem{Simonis:2015vja}
J.~Simonis, K.~Hebeler, J.D. Holt, J.~Menendez, A.~Schwenk, Phys. Rev. C
  \textbf{93}(1), 011302 (2016).
\newblock \doi{10.1103/PhysRevC.93.011302}

\bibitem{Drischler:2017wtt}
C.~Drischler, K.~Hebeler, A.~Schwenk, Phys. Rev. Lett. \textbf{122}(4), 042501
  (2019).
\newblock \doi{10.1103/PhysRevLett.122.042501}

\bibitem{Gezerlis:2013}
A.~Gezerlis, I.~Tews, E.~Epelbaum, S.~Gandolfi, K.~Hebeler, A.~Nogga,
  A.~Schwenk, Phys. Rev. Lett. \textbf{111}(3), 032501 (2013).
\newblock \doi{10.1103/PhysRevLett.111.032501}

\bibitem{Freunek:61123}
M.~Freunek, {N}ucleon-nucleon interaction in chiral effective field theory in
  configuration space.
\newblock Diplom (univ.), Univ. Bonn, Jülich (2007).
\newblock Record converted from VDB: 12.11.2012; Bonn, Univ., Dipl., 2007

\bibitem{Gezerlis:2014}
A.~Gezerlis, I.~Tews, E.~Epelbaum, M.~Freunek, S.~Gandolfi, K.~Hebeler,
  A.~Nogga, A.~Schwenk, Phys. Rev. C \textbf{90}(5), 054323 (2014).
\newblock \doi{10.1103/PhysRevC.90.054323}

\bibitem{Huth:2017wzw}
L.~Huth, I.~Tews, J.E. Lynn, A.~Schwenk, Phys. Rev. C \textbf{96}(5), 054003
  (2017).
\newblock \doi{10.1103/PhysRevC.96.054003}

\bibitem{Piarulli:2019cqu}
M.~Piarulli, I.~Tews, Front. in Phys. \textbf{7}, 245 (2020).
\newblock \doi{10.3389/fphy.2019.00245}

\bibitem{Tews:2015ufa}
I.~Tews, S.~Gandolfi, A.~Gezerlis, A.~Schwenk, Phys. Rev. C \textbf{93}(2),
  024305 (2016).
\newblock \doi{10.1103/PhysRevC.93.024305}

\bibitem{Piarulli:2015}
M.~Piarulli, L.~Girlanda, R.~Schiavilla, R.N. P\'erez, J.E. Amaro, E.R.
  Arriola, Phys. Rev. C \textbf{91}, 024003 (2015).
\newblock \doi{10.1103/PhysRevC.91.024003}

\bibitem{Piarulli:2016}
M.~Piarulli, L.~Girlanda, R.~Schiavilla, A.~Kievsky, A.~Lovato, L.E. Marcucci,
  S.C. Pieper, M.~Viviani, R.B. Wiringa, Phys. Rev. C \textbf{94}, 054007
  (2016).
\newblock \doi{10.1103/PhysRevC.94.054007}

\bibitem{Piarulli:2018prl}
M.~Piarulli, A.~Baroni, L.~Girlanda, A.~Kievsky, A.~Lovato, E.~Lusk, L.E.
  Marcucci, S.C. Pieper, R.~Schiavilla, M.~Viviani, R.B. Wiringa, Phys. Rev.
  Lett. \textbf{120}, 052503 (2018).
\newblock \doi{10.1103/PhysRevLett.120.052503}

\bibitem{Piarulli:2018}
A.~Baroni, R.~Schiavilla, L.E. Marcucci, L.~Girlanda, A.~Kievsky, A.~Lovato,
  S.~Pastore, M.~Piarulli, S.C. Pieper, M.~Viviani, R.B. Wiringa, Phys. Rev. C
  \textbf{98}, 044003 (2018).
\newblock \doi{10.1103/PhysRevC.98.044003}

\bibitem{King:2020}
G.B. King, L.~Andreoli, S.~Pastore, M.~Piarulli, R.~Schiavilla, R.B. Wiringa,
  J.~Carlson, S.~Gandolfi, Physical Review C \textbf{102}(2) (2020).
\newblock \doi{10.1103/physrevc.102.025501}

\bibitem{Schiavilla:2021}
R.~Schiavilla, L.~Girlanda, A.~Gnech, A.~Kievsky, A.~Lovato, L.E. Marcucci,
  M.~Piarulli, M.~Viviani, Physical Review C \textbf{103}(5) (2021).
\newblock \doi{10.1103/physrevc.103.054003}

\bibitem{Tews:2018sbi}
I.~Tews, L.~Huth, A.~Schwenk, Phys. Rev. C \textbf{98}(2), 024001 (2018).
\newblock \doi{10.1103/PhysRevC.98.024001}

\bibitem{Gandolfi:2020}
S.~Gandolfi, D.~Lonardoni, A.~Lovato, M.~Piarulli, Frontiers in Physics
  \textbf{8}, 117 (2020).
\newblock \doi{10.3389/fphy.2020.00117}

\bibitem{Hen:2017}
O.~Hen, G.A. Miller, E.~Piasetzky, L.B. Weinstein, Rev. Mod. Phys. \textbf{89},
  045002 (2017).
\newblock \doi{10.1103/RevModPhys.89.045002}

\bibitem{Abbott:2018}
B.P. Abbott, et~al., Phys. Rev. Lett. \textbf{121}, 161101 (2018).
\newblock \doi{10.1103/PhysRevLett.121.161101}

\bibitem{Li:2019}
B.A. Li, P.G. Krastev, D.H. Wen, N.B. Zhang, Eur. Phys. J. A \textbf{55}, 117
  (2019).
\newblock \doi{10.1140/epja/i2019-12780-8}

\bibitem{Dietrich:2020efo}
T.~Dietrich, M.W. Coughlin, P.T.H. Pang, M.~Bulla, J.~Heinzel, L.~Issa,
  I.~Tews, S.~Antier, Science \textbf{370}(6523), 1450 (2020).
\newblock \doi{10.1126/science.abb4317}

\bibitem{Tews:2018iwm}
I.~Tews, J.~Margueron, S.~Reddy, Phys. Rev. C \textbf{98}(4), 045804 (2018).
\newblock \doi{10.1103/PhysRevC.98.045804}

\bibitem{Demorest:2010bx}
P.~Demorest, T.~Pennucci, S.~Ransom, M.~Roberts, J.~Hessels, Nature
  \textbf{467}, 1081 (2010).
\newblock \doi{10.1038/nature09466}

\bibitem{Antoniadis:2013pzd}
J.~Antoniadis, P.C. Freire, N.~Wex, T.M. Tauris, R.S. Lynch, et~al., Science
  \textbf{340}, 6131 (2013).
\newblock \doi{10.1126/science.1233232}

\bibitem{Arzoumanian:2017puf}
Z.~Arzoumanian, et~al., Astrophys. J. Suppl. \textbf{235}(2), 37 (2018).
\newblock \doi{10.3847/1538-4365/aab5b0}

\bibitem{Cromartie:2019kug}
H.T. Cromartie, et~al., Nature Astron. \textbf{4}(1), 72 (2019).
\newblock \doi{10.1038/s41550-019-0880-2}

\bibitem{Miller:2019cac}
M.C. Miller, et~al., Astrophys. J. Lett. \textbf{887}, L24 (2019).
\newblock \doi{10.3847/2041-8213/ab50c5}

\bibitem{Riley:2019yda}
T.E. Riley, et~al., Astrophys. J. Lett. \textbf{887}, L21 (2019).
\newblock \doi{10.3847/2041-8213/ab481c}

\bibitem{LIGOScientific:2020aai}
B.P. Abbott, et~al., Astrophys. J. Lett. \textbf{892}(1), L3 (2020).
\newblock \doi{10.3847/2041-8213/ab75f5}

\bibitem{GBM:2017lvd}
B.~Abbott, et~al., Astrophys. J. Lett. \textbf{848}(2), L12 (2017).
\newblock \doi{10.3847/2041-8213/aa91c9}

\bibitem{Arcavi:2017xiz}
I.~Arcavi, et~al., Nature \textbf{551}, 64 (2017).
\newblock \doi{10.1038/nature24291}

\bibitem{Coulter:2017wya}
D.A. Coulter, et~al., Science  (2017).
\newblock \doi{10.1126/science.aap9811}.
\newblock [Science358,1556(2017)]

\bibitem{Rezzolla:2017aly}
L.~Rezzolla, E.R. Most, L.R. Weih, Astrophys. J. \textbf{852}(2), L25 (2018).
\newblock \doi{10.3847/2041-8213/aaa401}

\bibitem{Tews:2020ylw}
I.~Tews, P.T.H. Pang, T.~Dietrich, M.W. Coughlin, S.~Antier, M.~Bulla,
  J.~Heinzel, L.~Issa, Astrophys. J. Lett. \textbf{908}(1), L1 (2021).
\newblock \doi{10.3847/2041-8213/abdaae}

\bibitem{Al-Mamun:2020vzu}
M.~Al-Mamun, A.W. Steiner, J.~N\"attil\"a, J.~Lange, R.~O'Shaughnessy, I.~Tews,
  S.~Gandolfi, C.~Heinke, S.~Han, Phys. Rev. Lett. \textbf{126}(6), 061101
  (2021).
\newblock \doi{10.1103/PhysRevLett.126.061101}

\bibitem{Pang:2021jta}
P.T.H. Pang, I.~Tews, M.W. Coughlin, M.~Bulla, C.~Van Den~Broeck, T.~Dietrich,
  arXiv:2105.08688  (2021)

\bibitem{Huth:2021bsp}
S.~Huth, et~al., arXiv:2107.06229  (2021)

\bibitem{Essick:2021kjb}
R.~Essick, I.~Tews, P.~Landry, A.~Schwenk, arXiv:2102.10074  (2021)

\bibitem{Essick:2021ezp}
R.~Essick, P.~Landry, A.~Schwenk, I.~Tews, arXiv:2107.05528  (2021)

\bibitem{PREXII}
D.~Adhikari, H.~Albataineh, D.~Androic, K.~Aniol, D.S. Armstrong, T.~Averett,
  C.~Ayerbe~Gayoso, S.~Barcus, V.~Bellini, R.S. Beminiwattha, et~al., Phys.
  Rev. Lett. \textbf{126}(17), 172502 (2021).
\newblock \doi{10.1103/PhysRevLett.126.172502}

\bibitem{Reed:2021nqk}
B.T. Reed, F.J. Fattoyev, C.J. Horowitz, J.~Piekarewicz, Phys. Rev. Lett.
  \textbf{126}(17), 172503 (2021).
\newblock \doi{10.1103/PhysRevLett.126.172503}

\bibitem{Essick:2020flb}
R.~Essick, I.~Tews, P.~Landry, S.~Reddy, D.E. Holz, Phys. Rev. C
  \textbf{102}(5), 055803 (2020).
\newblock \doi{10.1103/PhysRevC.102.055803}

\bibitem{Most:2018hfd}
E.R. Most, L.R. Weih, L.~Rezzolla, J.~Schaffner-Bielich, Phys. Rev. Lett.
  \textbf{120}(26), 261103 (2018).
\newblock \doi{10.1103/PhysRevLett.120.261103}

\bibitem{Raaijmakers:2019qny}
G.~Raaijmakers, et~al., Astrophys. J. Lett. \textbf{887}(1), L22 (2019).
\newblock \doi{10.3847/2041-8213/ab451a}

\bibitem{Capano:2019eae}
C.D. Capano, I.~Tews, S.M. Brown, B.~Margalit, S.~De, S.~Kumar, D.A. Brown,
  B.~Krishnan, S.~Reddy, Nature Astron. \textbf{4}(6), 625 (2020).
\newblock \doi{10.1038/s41550-020-1014-6}

\bibitem{Entem:2017gor}
D.R. Entem, R.~Machleidt, Y.~Nosyk, Phys. Rev. C \textbf{96}(2), 024004 (2017).
\newblock \doi{10.1103/PhysRevC.96.024004}

\bibitem{Reinert:2017usi}
P.~Reinert, H.~Krebs, E.~Epelbaum, Eur. Phys. J. A \textbf{54}(5), 86 (2018).
\newblock \doi{10.1140/epja/i2018-12516-4}

\end{thebibliography}

\end{document}